\documentclass[a4paper]{article}

\usepackage{INTERSPEECH2019}

\usepackage{scalerel}
\usepackage[justification=centering]{caption} 
\usepackage{multirow, makecell} 

\title{Investigation of F0 conditioning and Fully Convolutional Networks in Variational Autoencoder based Voice Conversion}
\name{ \begin{tabular}{c}
		Wen-Chin Huang$^{12}$, Yi-Chiao Wu$^2$, Chen-Chou Lo$^1$, Patrick Lumban Tobing$^2$, \\
		Tomoki Hayashi$^2$, Kazuhiro Kobayashi$^2$, Tomoki Toda$^2$, Yu Tsao$^1$, Hsin-Min Wang$^1$, 
	\end{tabular}
}
\address{$^{1}$ Academia Sinica, Taiwan \\
	$^{2}$ Nagoya University, Japan
}

\email{wen.chinhuang@g.sp.m.is.nagoya-u.ac.jp}

\begin{document}

\maketitle
\begin{abstract}
	In this work, we investigate the effectiveness of two techniques for improving variational autoencoder (VAE) based voice conversion (VC). First, we reconsider the relationship between vocoder features extracted using the high quality vocoders adopted in conventional VC systems, and hypothesize that the spectral features are in fact F0 dependent. Such hypothesis implies that during the conversion phase, the latent codes and the converted features in VAE based VC are in fact source F0 dependent. To this end, we propose to utilize the F0 as an additional input of the decoder. The model can learn to disentangle the latent code from the F0 and thus generates converted F0 dependent converted features. Second, to better capture temporal dependencies of the spectral features and the F0 pattern, we replace the frame wise conversion structure in the original VAE based VC framework with a fully convolutional network structure. Our experiments demonstrate that the degree of disentanglement as well as the naturalness of the converted speech are indeed improved.
\end{abstract}

\section{Introduction}

Voice conversion (VC) aims to convert the speech from a source to that of a target without changing the linguistic content. Numerous approaches have been proposed, such as Gaussian mixture model (GMM)-based methods \cite{VC,GMM-VC}, deep neural network (DNN)-based methods \cite{ANN-VC, layerwise-VC}, and exemplar-based methods \cite{exemplar-noisy-VC,exemplar-residual-VC,LLE-VC}. Most of them require parallel training data, i.e., the source and target speakers utter the same transcripts for training. Since such data is hard to collect, non-parallel training has long remained one of the ultimate goals in VC. 

Recently, VAEs \cite{VAE} have been successfully applied to VC \cite{VAE-VC}, which we will refer to as VAE-VC. Specifically, the spectral conversion function is composed of an encoder-decoder pair. The encoder first encodes the input spectral feature into a latent code. Then, the decoder mixes the latent code and the target speaker code to generate the output. The encoder-decoder network and the speaker codes are trained by back-propagation of the reconstruction error, along with a Kullback-Leibler (KL)-divergence loss that regularizes the distribution of the latent variable, thus there is no need for parallel training data. The success of this framework implies that the encoder learns to eliminate the speaker dependent information from the input, making the latent code speaker independent.

\begin{figure*}[t]
  \centering
  \includegraphics[width=0.75\textwidth]{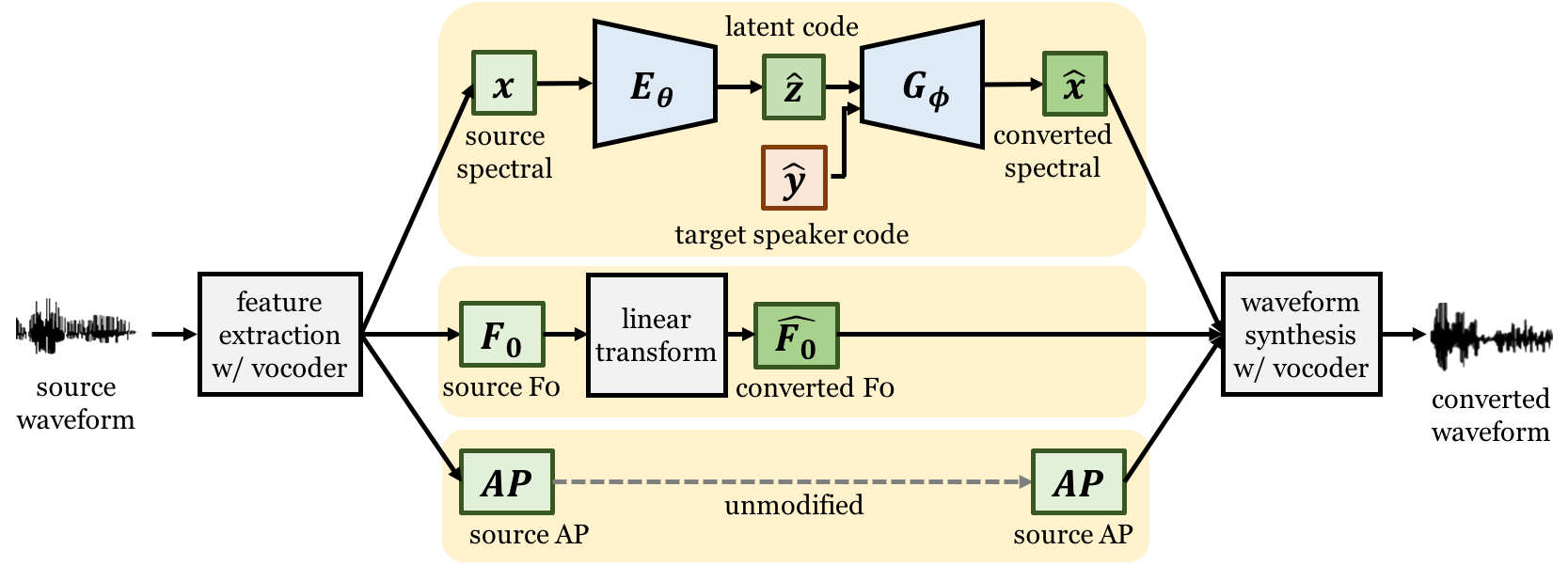}
  \centering
  \captionof{figure}{Illustration of the VAE-VC framework. Following traditional VC systems, a vocoder first parameterizes the waveform into acoustic features, which are then converted in different streams, and finally the converted features are used to synthesize the converted waveform by a vocoder.}
  \label{fig:vae-vc}
\end{figure*}

Following conventional VC systems \cite{NU-VCC2016}, the VAE-VC framework first utilizes high quality vocoders such as WORLD \cite{WORLD} and STRAIGHT \cite{STRAIGHT} to extract different kinds of acoustic features, e.g., spectral feature and fundamental frequency (F0). As depicted in Fig.~\ref{fig:vae-vc}, these features are then converted separately, and a waveform synthesizer finally generates the converted waveform using the converted features. The validity of converting acoustic features in different feature streams comes from the assumption that these features are independent from each other. However, during feature extraction, both WORLD and STRAIGHT extract the F0 first, then use the extracted F0 to obtain the spectral feature. Thus, we hypothesize that the spectral features are in fact F0 dependent.

Based on this hypothesis, we may imply that during the conversion phase in VAE-VC, since the encoder was never trained to eliminate F0 information, the latent code extracted from the source spectral feature still contains information of the source F0, thus the converted feature is source F0 dependent. It can be assumed that the conversion performance can suffer from this flaw. In other words, if the converted spectral feature can be made converted F0 dependent, the performance will improve. This is analogous to a previous work that models the cross stream dependency in Hidden Markov Model based speech synthesis \cite{CSDM-HMM-SPSS,CSDM-F0-HMM-SPSS}

\begin{figure}[t]
  \centering
  \includegraphics[width=0.45\textwidth]{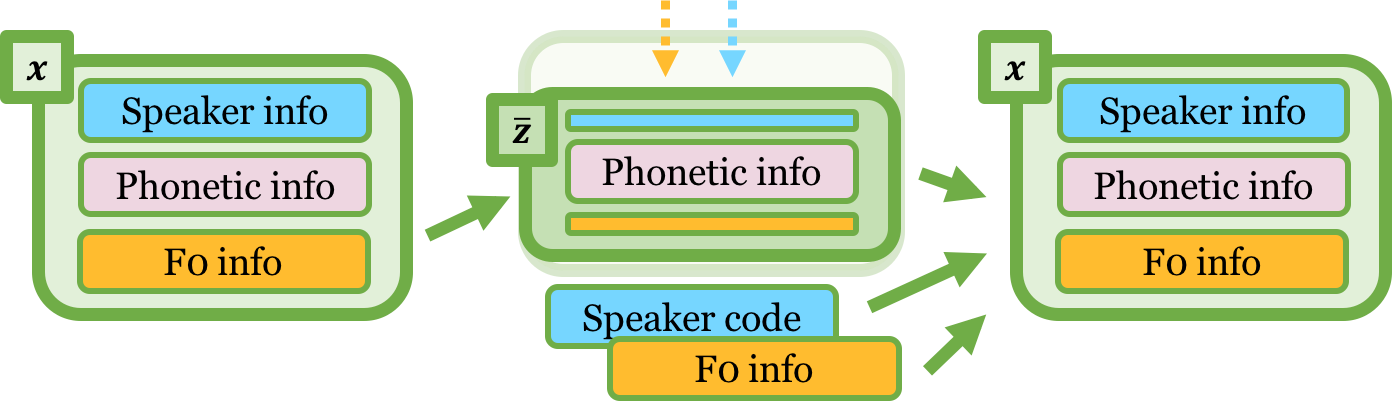}
  \centering
  \captionof{figure}{Disentangling the latent code with condition variables. By providing the speaker code and F0 explicitly, the encoder learns to discard as much speaker and F0 information as possible, thereby generating compact latent representations.}
  \label{fig:disentangle}
\end{figure}

How do we obtain converted spectral features that are converted F0 dependent? In VAE-VC, by conditioning the decoder with the speaker code, the encoder can learn to eliminate speaker dependent information from the input, thus make the latent code speaker independent. We may assume that, similarly, by conditioning the decoder with F0, we may disentangle the latent code from F0, as illustrated in Fig.~\ref{fig:disentangle}. As a result, during the conversion phase, given the converted F0, we may obtain the desired converted F0 dependent converted spectral feature.

However, applying the above mentioned concept to VAE-VC may be somehow problematic. The original VAE-VC performs conversion in a frame wise manner, i.e., no temporal relationship is considered. Under such structure, if the model is given an F0 value per frame, it is concerned whether the encoder can actually learn to eliminate F0 information effectively. Therefore, we assume that by designing the network to be able to acquire an input sequence, the model can benefit from capturing the F0 contour, and thus better disentangle the latent code.

In this work, we investigate two techniques to improve the general VAE-VC framework. Specifically, we condition the decoder with F0, and adopt the fully convolutional network (FCN) \cite{FCN} to consider temporal dependencies of the inputs. Our contributions are:
	\begin{itemize}
	  \item We reconsidered the relationship between different vocoder features, and hypothesized that the use of F0 as an additional condition variable of the model can eliminate the source F0 information in the encoded latent codes, and as a result obtain the converted F0 dependent converted spectral feature. We verified this hypothesis by showing that the latent code obtained in this way is indeed more F0 independent through objective measures, and evaluated the conversion performance.
	  \item We adopted the FCN structure so that the model can take an input sequence in order to consider the temporal relationship of the spectral features and F0 pattern. Note that the FCN structure was first combined with VAE-VC in \cite{ACVAE-VC}, but the impact of FCN was not solely examined. We provide detailed experiments to examine the effectiveness of this structure.
	\end{itemize}

\section{Related work on VAE-VC}

\subsection{VAE-VC}

Figure~\ref{fig:vae-vc} illustrates a VAE-VC system \cite{VAE-VC}. The core of VAE-VC is an encoder-decoder network, which models the WORLD spectra (SP). During training, given an input spectral frame $\vec{x}$, the encoder $E_\theta$ with parameter set $\theta$ encodes $\vec{x}$ into a latent code: $\vec{z}=E_\theta(\vec{x})$. The speaker code $\vec{y}$ of the input frame, along with $\vec{z}$, are passed to the decoder $G_\phi$ with parameter set $\phi$ to reconstruct the input. This reconstruction process can be written as:
\begin{equation}
	\label{eq:self-recon}
	\bar{\vec{x}}=G_\phi(\vec{z},\vec{y})=G_\phi(E_\theta(\vec{x}),\vec{y}).
\end{equation}
The model parameters can be obtained by maximizing the variational lower bound:
\begin{equation}
\label{eq:vae_loss}
	\mathcal{L}_{vae}(\theta,\phi;\vec{x}, \vec{y}) = \mathcal{L}_{recon}(\vec{x},\vec{y})+\mathcal{L}_{lat}(\vec{x}),
\end{equation}
\begin{equation}
	\label{eq:recon_loss}
	\mathcal{L}_{recon}(\vec{x},\vec{y})=\mathbb{E}_{\vec{z}\sim q_\theta(\vec{z}|\vec{x})}\bigl[\log p_\phi(\vec{x}|\vec{z},\vec{y})\bigr],
\end{equation}
\begin{equation}
	\label{eq:lat_loss}
	\mathcal{L}_{lat}(\vec{x})=-D_{KL}(q_\theta(\vec{z}|\vec{x}) \Vert p(\vec{z})),
\end{equation}
where $q_\theta(\vec{z}|\vec{x})$ is the approximate posterior, $p_\phi(\vec{x}|\vec{z},\vec{y})$ is the data likelihood, and $p(\vec{z})$ is the prior distribution of the latent space. $\mathcal{L}_{recon}$ is simply a reconstruction term as in any vanilla auto encoder, whereas $\mathcal{L}_{lat}$ regularizes the encoder to align the approximate posterior with the prior distribution.

In the conversion phase, one could use \eqref{eq:self-recon} to formulate the conversion function $f$ with the target speaker $\hat{\vec{y}}$:
\begin{equation}
	\label{eq:conversion}
	\hat{\vec{x}}=f(\vec{x}, \hat{\vec{y}})=G_\phi(\vec{z},\hat{\vec{y}})=G_\phi(E_\theta(\vec{x}),\hat{\vec{y}}).
\end{equation}

There is a line of work extending the VAE-VC framework. \cite{VAE-GAN-VC} was the first VC framework to incorporate generative adversarial network (GAN) to improve spectral modeling. \cite{VAE-PPG-DVEC-VC} utilized external modules such as automatic speech recognition and speaker verification systems to obtain phonetic posteriorgrams and d-vectors to improve the performance. \cite{ACVAE-VC} borrowed the idea of auxiliary classifiers from conditional image generation to force the decoder to preserve more speaker characteristics, and further used an FCN structure to take sequential input features.

In the following subsection, we will introduce the baseline system we use in this paper.

\subsection{CDVAE-VC}

\cite{CDVAE} proposed a cross-domain VAE framework, by extending the conventional VAE framework to jointly consider two kinds of spectral features, SPs and Mel-Cepstral Coefficients (MCCs), extracted from the same observed speech frame, to utilize their different properties. This framework, which we will refer to as CDVAE-VC, is a collection of encoder-decoder pairs, one for each kind of spectral feature. Considering the SPs and MCCs as two kinds of spectral features (denoted as $\vec{x}_{\scaleto{SP}{3pt}}$ and $\vec{x}_{\scaleto{MCC}{3pt}}$ ), we define the following losses:
\begin{align}
	\vec{z}_{\scaleto{SP}{3pt}}&=E_{\scaleto{SP}{3pt}}(\vec{x}_{\scaleto{SP}{3pt}}), \vec{z}_{\scaleto{MCC}{3pt}}=E_{\scaleto{MCC}{3pt}}(\vec{x}_{\scaleto{MCC}{3pt}}),\\
	\bar{\vec{x}}_{\scaleto{S-S}{3pt}}&=G_{\scaleto{SP}{3pt}}(\vec{z}_{\scaleto{SP}{3pt}},\vec{y}), \bar{\vec{x}}_{\scaleto{M-M}{3pt}}=G_{\scaleto{MCC}{3pt}}(\vec{z}_{\scaleto{MCC}{3pt}},\vec{y}),\\
	\bar{\vec{x}}_{\scaleto{S-M}{3pt}}&=G_{\scaleto{MCC}{3pt}}(\vec{z}_{\scaleto{SP}{3pt}},\vec{y}), \bar{\vec{x}}_{\scaleto{M-S}{3pt}}=G_{\scaleto{SP}{3pt}}(\vec{z}_{\scaleto{MCC}{3pt}},\vec{y}),\\
	\mathcal{L}_{in}&=\mathcal{L}_{recon}(\bar{\vec{x}}_{\scaleto{S-S}{3pt}},\vec{y})+\mathcal{L}_{recon}(\bar{\vec{x}}_{\scaleto{M-M}{3pt}},\vec{y}),\\
	\mathcal{L}_{KLD}&=\mathcal{L}_{lat}(\vec{x}_{\scaleto{SP}{3pt}})+\mathcal{L}_{lat}(\vec{x}_{\scaleto{MCC}{3pt}}),\\
	\mathcal{L}_{cross}&=\mathcal{L}_{recon}(\bar{\vec{x}}_{\scaleto{S-M}{3pt}},\vec{y})+\mathcal{L}_{recon}(\bar{\vec{x}}_{\scaleto{M-S}{3pt}},\vec{y}),\\
	\mathcal{L}_{sim}&=\Vert \vec{z}_{\scaleto{SP}{3pt}}-\vec{z}_{\scaleto{MCC}{3pt}} \Vert _1 ,
\end{align}
where $E_{\scaleto{SP}{3pt}}$ and $E_{\scaleto{MCC}{3pt}}$ are the encoders for SP and MCC, respectively, while  $G_{\scaleto{SP}{3pt}}$ and $G_{\scaleto{MCC}{3pt}}$ are decoders.

In short, two extra reconstruction streams were introduced. By optimizing the cross-domain reconstruction loss, $\vec{z}_{\scaleto{SP}{3pt}}$ is enforced to contain enough information to reconstruct $\vec{x}_{\scaleto{MCC}{3pt}}$, and vice versa. As a result, the behavior of the encoders from both feature domains are constrained to be the same, i.e., they are expected to extract similar latent information from different types of input spectral features. To explicitly reinforce this constraint, a latent similarity loss was also included.

The final objective is as follows:
\begin{equation}
	\label{eq:cdvae_objective}
	\mathcal{L}_{cdvae}=\mathcal{L}_{in}+\mathcal{L}_{KLD}+\mathcal{L}_{cross}+\mathcal{L}_{sim}.
\end{equation}

The model parameters can be learned by minimizing \eqref{eq:cdvae_objective}. In the conversion phase, there are four conversion paths (i.e., two within-domain and two cross-domain paths). As reported in \cite{CDVAE}, the MCC-MCC path gave the best performance in terms of subjective measure, which matched the common assumption that MCCs are related to human perception.

\subsection{Problem definition}
\label{flaws}

Here we once again point out the flaws of the general VAE-VC framework. First, since the model is built on a frame wise basis, the network is limited in modeling the temporal dependencies of speech. Second, it is possible that the spectral features extracted using a vocoder is actually F0 dependent. Thus, F0 information in the source spectral feature might remain in the encoded latent code as well as the converted spectral feature, thereby damaging the conversion performance.

\begin{figure}[t]
  \centering
  \includegraphics[width=0.5\textwidth]{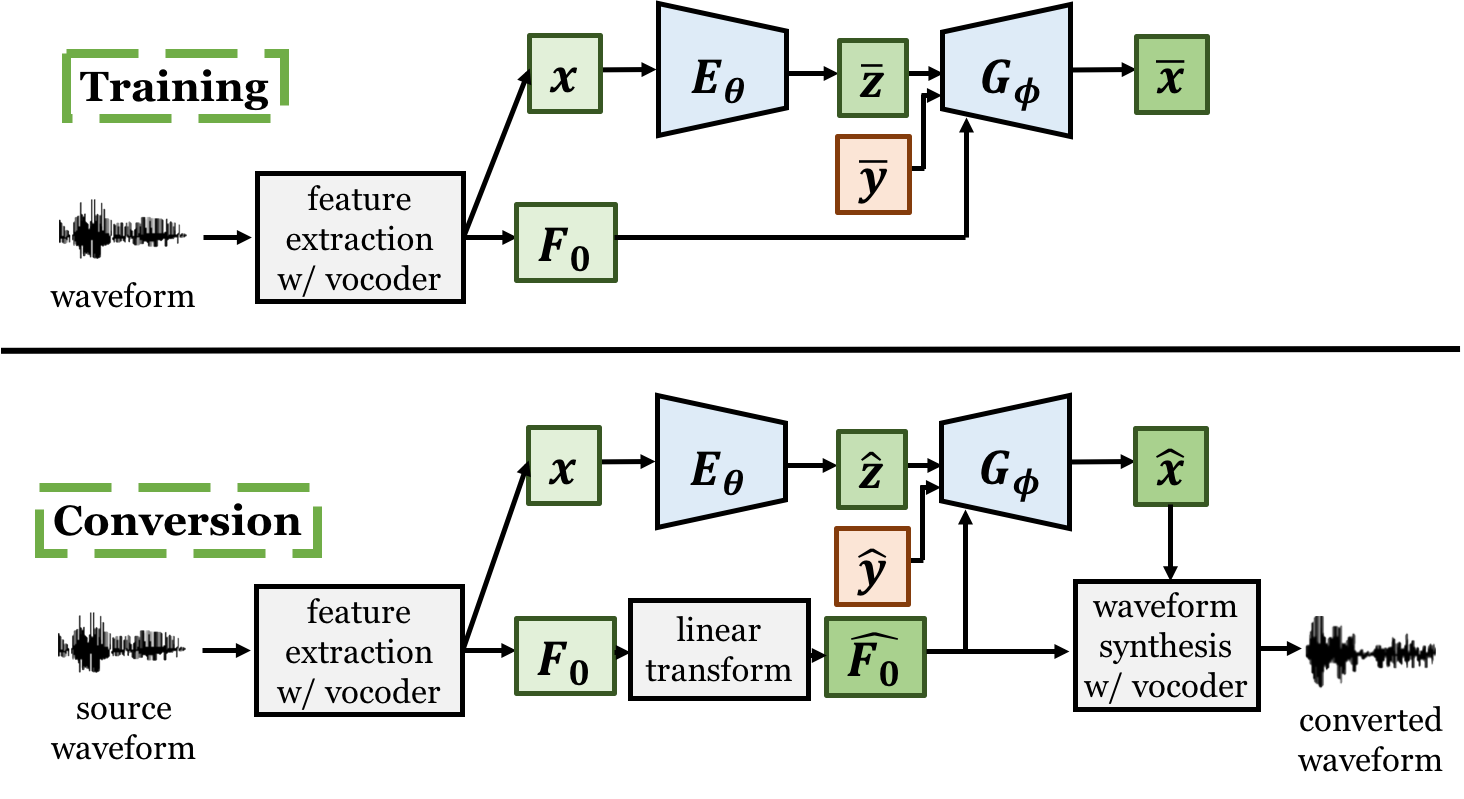}
  \centering
  \captionof{figure}{The proposed framework with F0 conditioning.}
  \label{fig:f0vae}
\end{figure}

\section{Investigated Methods}

In this section, we examine two mechanisms to overcome the disadvantages in VAE-VC mentioned in Section~\ref{flaws}.

\subsection{Modeling time dependencies with the FCN structure}

When it comes to sequential models, the recurrent neural network (RNN) is a commonly chosen network structure. Nonetheless, we follow \cite{ACVAE-VC} and adopt the FCN structure. There are several reasons why we choose FCNs over RNNs. First, the nature of RNNs introduces high computational costs, and convolutional layers make parallel computation feasible. Second, RNNs have an infinitely large receptive field in theory, and we think this is unnecessary in our task. In contrast, by adjusting the depth and kernel sizes, convolutional neural networks can be flexibly designed to have a finite, reasonably large receptive field. Note that here the output of our model is still of the same length as the input. Although sequence to sequence based models, which can generate output sequences of variable length, have been successfully applied to VC \cite{S2S-VC,S2S-GAN-VC,S2S-iFLYTEK-VC,ATT-S2S-VC, CONV-S2S-VC}, we will show that only considering temporal dependencies can bring significant improvements to VAE-VC.

\subsection{Conditioning on F0}

We propose to use F0 as an additional condition variable in order to eliminate the F0 information in the latent code, as shown in Fig.~\ref{fig:f0vae}. Specifically, during training, given the F0 contour of the input $F_0$, we modify~\eqref{eq:self-recon} as:
\begin{equation}
	\label{eq:f0-self-recon}
	\bar{\vec{x}}=G_\phi(\vec{z},\vec{y}, F_0)=G_\phi(E_\theta(\vec{x}),\vec{y}, F_0).
\end{equation}
During conversion, given the converted F0 contour $\hat{F_0}$, we modify~\eqref{eq:f0-self-recon} to obtain:
\begin{equation}
	\label{eq:conversion}
	\hat{\vec{x}}=f(\vec{x}, \hat{\vec{y}}, \hat{F_0})=G_\phi(\vec{z},\hat{\vec{y}}, \hat{F_0})=G_\phi(E_\theta(\vec{x}),\hat{\vec{y}}, \hat{F_0}).
\end{equation}
In our preliminary experiments, we tested several combinations of prosodic features, including \textit{F0}, \textit{continuous (interpolated, cont) F0 + uv symbol} and \textit{cont F0 + uv + band APs}, and found that \textit{cont F0 + uv} had the best performance. 

\begin{figure}[t]
  \centering
  \includegraphics[width=0.45\textwidth]{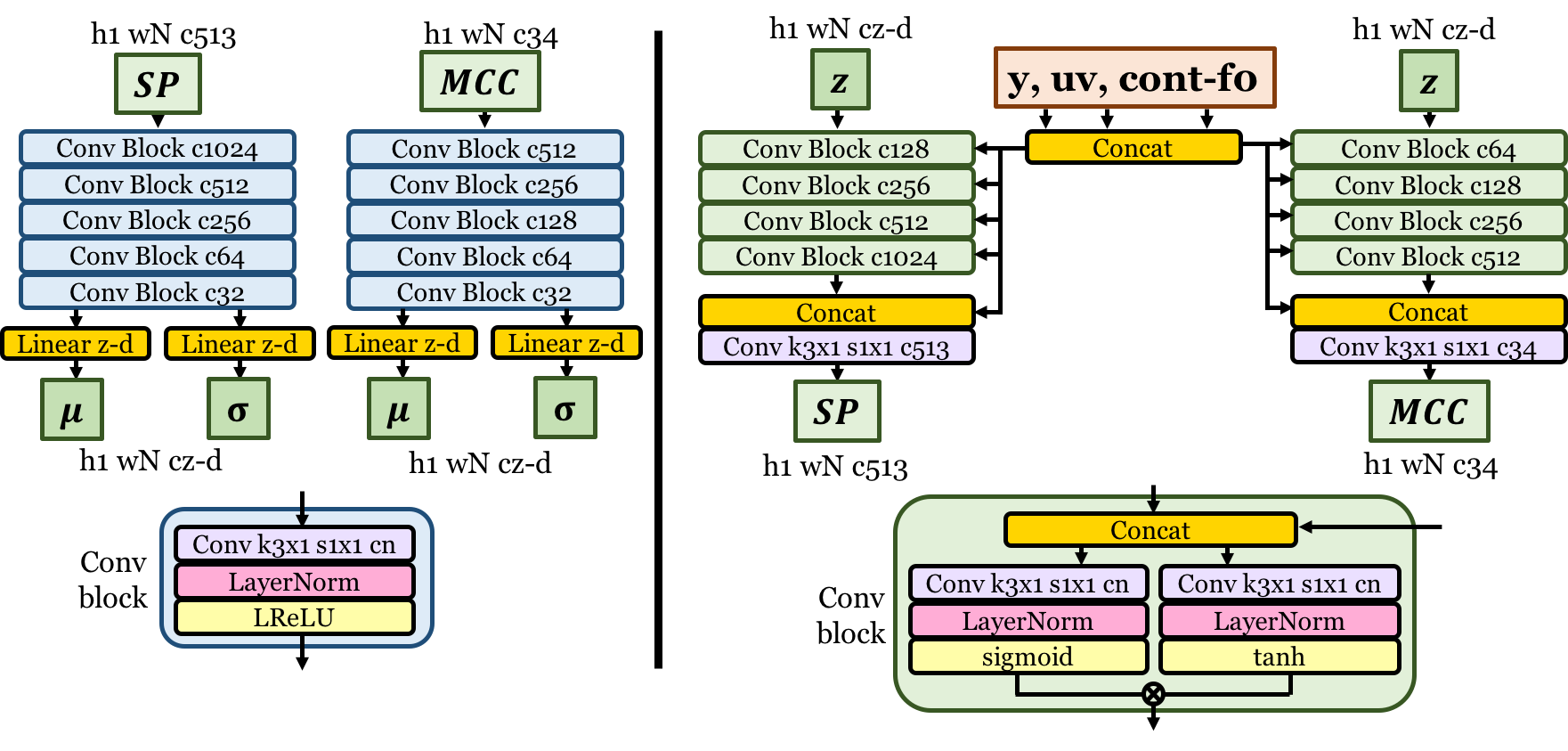}
  \centering
  \captionof{figure}{Model architecture. The input is of length N. h, w, c means height, width and channels. z-d means latent code dimension. LReLU, LayerNorm means leaky rectified activation function and layer normalization layer. k, s, c means kernel size, stride and output channels. }
  \label{fig:arch}
\end{figure}

\section{Experimental Evaluation}

\subsection{Experimental settings}
 
We evaluated our proposed methods on the Voice Conversion Challenge 2018 dataset \cite{vcc2018}, which included recordings of professional US English speakers with a sampling rate of 22050 Hz. The dataset consisted of 81/35 utterances per speaker for training/testing sets, respectively. We used the first 70 utterances of the training set of all speakers for training, the remaining 11 for validation, and the 35 in the testing set of speakers SF1, SF2, SM1, SM2, TF1, TF2, TM1, TM2 to form 16 conversion pairs for evaluation. The WORLD vocoder \cite{WORLD} was adopted to extract acoustic features including 513-dimensional SPs, 513-dimensional APs and F0. The SPs were normalized to unit-sum, and the normalizing factor was taken out and thus not modified. 35-dimensional MCCs were further extracted from the SPs. In the conversion phase, the energy and APs were kept unmodified, and the F0 was converted using a linear mean-variance transformation in the log domain. 

The baseline system was the CDVAE-VC \cite{CDVAE} system. On top of CDVAE, we first replace the original frame wise structure with FCNs to take sequential inputs. We will refer to this model as FCN-CDVAE. The architecture, as illustrated in Fig.~\ref{fig:arch}, was very similar to \cite{ACVAE-VC}, using gated linear units activation function and skip connections in the decoders to better propagate the conditional information. Following \cite{cyclegan-vc, CHOU-NPVC}, we randomly sampled 128 frames with overlap during training. For both frame wise and sequential models, the batch size, latent dimension and speaker code dimension were all 16, and the models were trained using the Adam optimizer \cite{ADAM} with learning rate, $\beta_1$ and $\beta_2$ set to 0.0001, 0.5, and 0.999. Note that the speaker codes were randomly initialized and optimized, as in \cite{CDVAE}.

We applied the F0 conditioning mechanism to the FCN-CDVAE model, referred to as F0-FCN-CDVAE. We concatenated these features with the speaker code in the feature axis.

\begin{table}[t]
	\centering
	\captionsetup{justification=centering}
	\caption{Mean Mel-cepstral distortion [dB] of all non-silent frames from the baseline and proposed models.}
	\centering
	\begin{tabular}{ l c c c c c }
		\toprule	
		\textbf{System} & \textbf{F-F} & \textbf{F-M} & \textbf{M-F} & \textbf{M-M} & \textbf{Avg.}\\
		\midrule
		\textbf{CDVAE} & 6.67 & 6.31 & 6.71 & 5.97 & 6.42\\
		\textbf{FCN-CDVAE} & 6.57 & 6.27 & 6.97 & 5.76 & 6.39\\
		\textbf{F0-FCN-CDVAE} & 6.56 & 6.31 & 6.86 & 5.79 & 6.38\\
	\end{tabular}
	\label{tab:MCD}
\end{table}


\begin{table}[t]
	\centering
	\captionsetup{justification=centering}
	\caption{MOS for naturalness with 95\% confidence intervals.}
	\centering
	\begin{tabular}{ l c }
		\toprule	
		\textbf{System} & \textbf{Avg.}\\
		\midrule
		\textbf{CDVAE} &  2.45 $\pm$ 0.13\\
		\textbf{FCN-CDVAE} & 2.89 $\pm$ 0.15\\
		\textbf{F0-FCN-CDVAE} & 2.84 $\pm$ 0.17\\
		\textbf{Target} & 4.96 $\pm$ 0.04\\
	\end{tabular}
	\label{tab:MOS}
\end{table}

\begin{table}[t]
	\centering
	\captionsetup{justification=centering}
	\caption{RMSE distance and cosine similarity of latent codes extracted from parallel sentences of source-target pairs over non-silent frames.}
	\centering
	\begin{tabular}{ l c c c c c }
		\toprule	
		\textbf{System} & \textbf{F-F} & \textbf{F-M} & \textbf{M-F} & \textbf{M-M} & \textbf{Avg.}\\
		\midrule
		\multicolumn{2}{l}{\textit{[RMSE]}} \\
		\textbf{FCN-CDVAE} & .335 & .337 & .353 & .304 & .333\\
		\textbf{F0-FCN-CDVAE} & .284 & .286 & .302 & .260 & .283\\
		\midrule
		\multicolumn{2}{l}{\textit{[Cosine Similarity]}} \\
		\textbf{FCN-CDVAE} & .530 & .502 & .475 & .588 & .524\\
		\textbf{F0-FCN-CDVAE} & .579 & .547 & .519 & .616 & .565\\
	\end{tabular}
	\label{tab:latent-distance}
\end{table}

\begin{figure}[t]

\begin{minipage}[b]{0.22\textwidth}
  \centering
  \includegraphics[width=\textwidth]{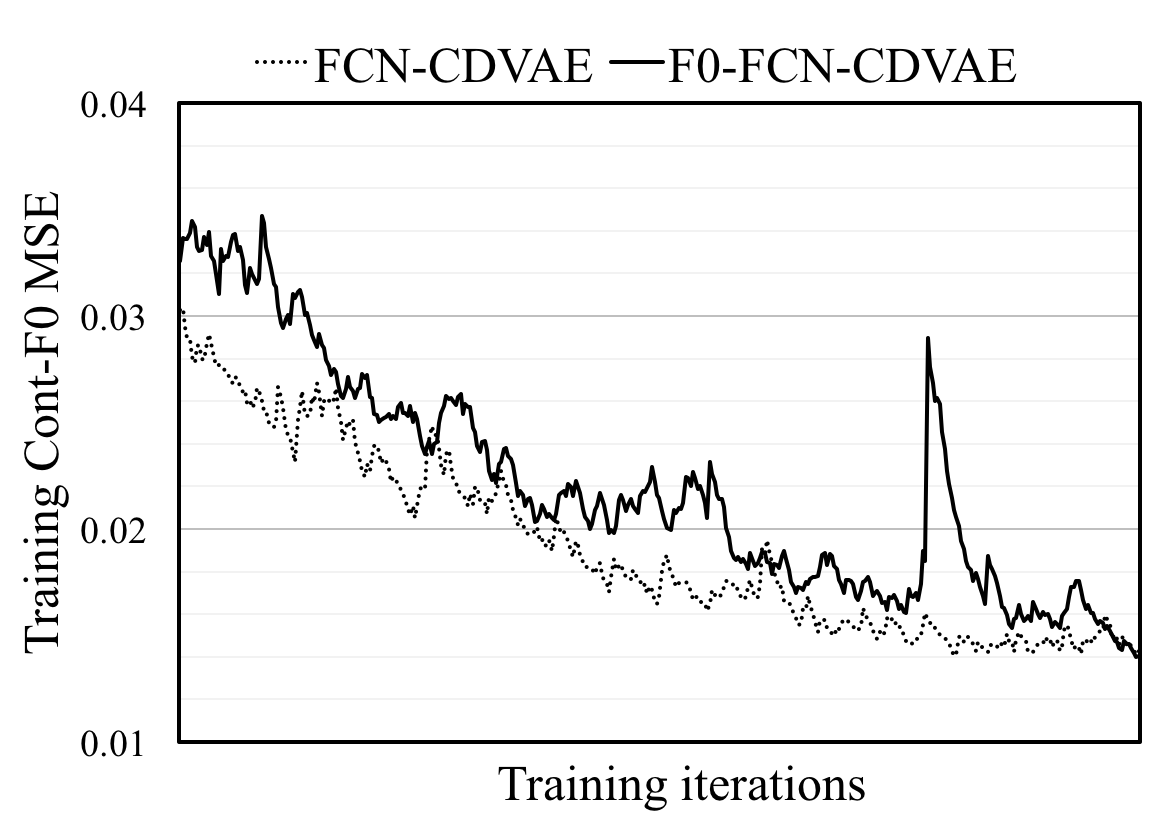}
  \captionof{figure}{Training Cont-F0 MSE of the F0 prediction network.}
  \label{fig:latent2f0-f0}
\end{minipage}
~
\begin{minipage}[b]{0.22\textwidth}
  \centering
  \includegraphics[width=\textwidth]{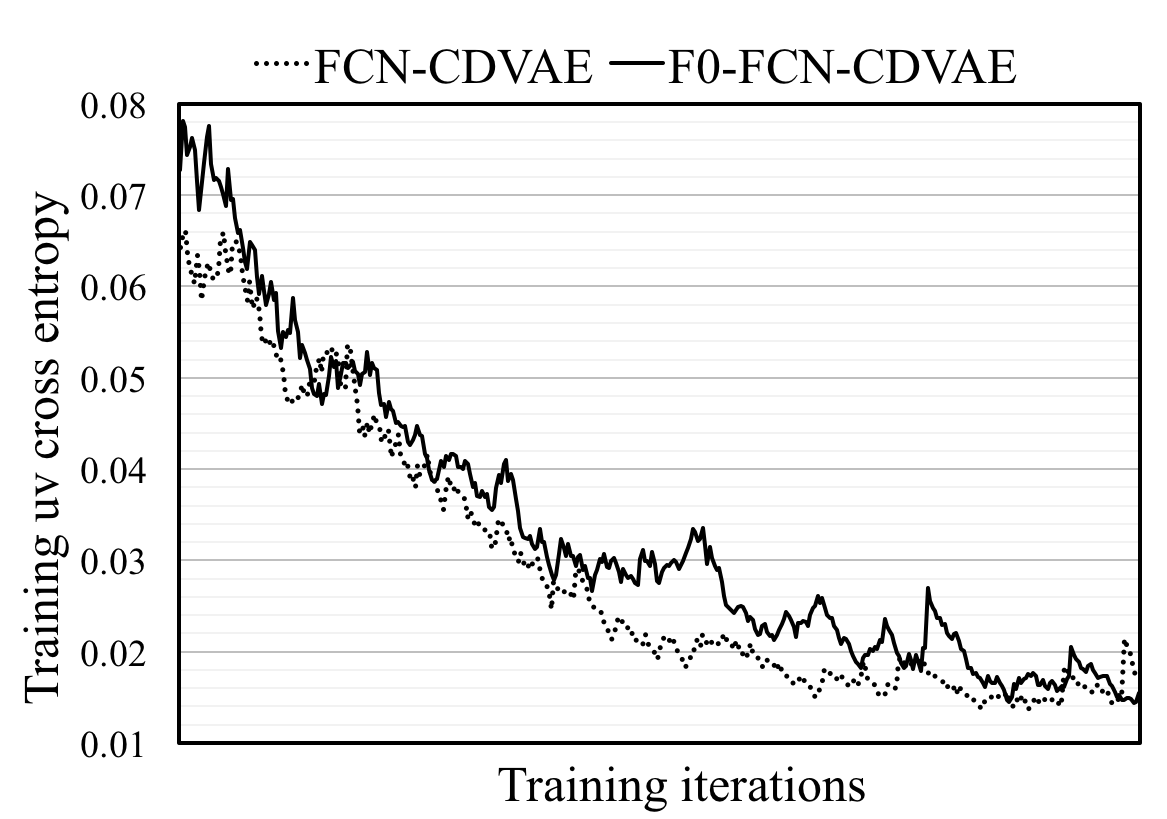}
  \captionof{figure}{Training uv cross entropy of the F0 prediction network.}
  \label{fig:latent2f0-uv}
\end{minipage}
\end{figure}

\subsection{Effectiveness of FCN}

We first examined the effectiveness of FCN. Table~\ref{tab:MCD} shows the mean Mel-cepstral distortion (MCD) values, and Table~\ref{tab:MOS} shows the mean opinion scores (MOS) obtained from a listening test where 10 participants were asked to evaluate the naturalness of the speech on a five-point scale, including the natural target speech. We may conclude from the results that the FCN structure indeed improved the overall performance in terms of both MCD and MOS, except for the M-F conversion pairs, where we will leave the investigation for future work. 

\subsection{Effect of F0 conditioning}
\label{ssec:exp-F0vae}

We then examined the F0 conditioning mechanism. We conducted several experiments to check if the latent codes are indeed more disentangled from F0. First, consider two sentences with the same content uttered by the source and the target speakers. An ideal encoder should extract identical latent codes from these two sentences since the phonetic contents are the same, though with different styles such as F0. We measure the distance between latent code pairs in terms of root mean squared error (RMSE) and cosine similarity, as in Table~\ref{tab:latent-distance}. The results show that the latent code pairs extracted using F0-FCN-CDVAE are closer in terms of RMSE and cosine similarity.

The amount of F0 information that resides in the latent code also reflects the degree of disentanglement. Following \cite{Representation-WNV}, we trained an F0 prediction network and reported the training loss. Specifically, a network with the same architecture as the encoder in Fig.~\ref{fig:arch} was trained to take a sequence of latent codes as input and predict the corresponding \textit{cont-F0 + uv}. We assumed that less F0 information left in the latent codes results in worse training performance. Fig.~\ref{fig:latent2f0-f0} and Fig.~\ref{fig:latent2f0-uv} show the training \textit{cont-F0} mean squared error (MSE) and \textit{uv} cross entropy. As expected, the training losses with latent codes extracted using F0-FCN-CDVAE were higher, suggesting less F0 information present in the latent codes.

We finally examined the performance of VC. As reported in Tables~\ref{tab:MCD} and~\ref{tab:MOS}, F0-FCN-CDVAE had an similar performance compared with FCN-CDVAE in terms of MCD and MOS.

\section{Discussions and Conclusions}

In this work, we investigated two approaches to improve VAE-VC. A FCN structure was applied to take sequential inputs rather than performing conversion frame by frame, thus capable of capturing the temporal relationship of speech. The F0 conditioning mechanism, motivated by a reconsideration of the relationship between vocoder features, helps eliminate residual F0 information in the latent code that might potentially harm the conversion performance. The experimental evaluations showed that the impact of FCNs on the objective measures and subjective speech naturalness assessment was positive. On the other hand, F0 conditioning showed promising results in increasing the degree of disentanglement of latent codes, and achieved high speech quality equivalent to FCN-CDVAE. Speech samples are available at \cite{F0-FCN-CDVAE-Demo}.
	
We attribute the insignificant improvement brought by the F0 conditioning scheme to a mismatch between training and conversion. The converted F0 obtained through such a simple F0 conversion process adopted in this and many past works is far from natural. As a result, in the conversion phase, the input combination which consisted of latent codes extracted from normal MCCs and the unnatural converted F0 might have not be seen by the model during training, thereby causing a degradation in quality.

While the above mentioned mismatch could be one possible reason, we would like to highlight that, although the motivation of applying F0 conditioning to VAE-VC was based on the assumption that the vocoder spectral features are F0 dependent, the design of vocoders was to \textit{separate} these two features as much as possible. The amount of F0 information that resides in the spectral features might already be small enough for our proposed mechanism to eliminate. In the future, we plan to apply this general idea to rawer input features that are richer in F0, e.g., magnitude spectrograms.\\
\noindent\textbf{Acknowledgements}: This work was partly supported by JSPS KAKENHI Grant Numbers JP17H06101 and 17H01763, as well as MOST-Taiwan Grants 105-2221-E001-012-MY3 and 107-2221-E-001-008-MY3.

\bibliographystyle{IEEEtran}

\bibliography{is2019}

\end{document}